\documentclass[prb,twocolumn,keyword,superscriptaddress]{revtex4-2}

\usepackage{graphicx}
\usepackage{dcolumn}
\usepackage{bm}
\usepackage{amssymb}
\usepackage{amsmath}
\usepackage{xcolor}
\usepackage{float}
\usepackage{multirow}
\usepackage{comment}
\usepackage[colorlinks=true,citecolor=blue]{hyperref}

\begin{document}

\title{Altermagnetic Flatband-Driven Fermi Surface Geometry for \\Giant Tunneling Magnetoresistance}

\author{Xingyue Yang}
\affiliation{State Key Laboratory for Mesoscopic Physics and School of Physics, Peking University, Beijing 100871, P. R. China}
\affiliation{Science, Mathematics and Technology (SMT) Cluster, Singapore University of Technology and Design, Singapore 487372, Republic of Singapore}

\author{Shibo Fang}
\affiliation{Science, Mathematics and Technology (SMT) Cluster, Singapore University of Technology and Design, Singapore 487372, Republic of Singapore}

\author{Zongmeng Yang}
\affiliation{State Key Laboratory for Mesoscopic Physics and School of Physics, Peking University, Beijing 100871, P. R. China}
\affiliation{Science, Mathematics and Technology (SMT) Cluster, Singapore University of Technology and Design, Singapore 487372, Republic of Singapore}

\author{Pin Ho}
\affiliation{Quantum Innovation Centre (Q.InC), Agency for Science Technology and Research (A*STAR), Singapore 138634, Republic of Singapore}

\author{Jing Lu}
\email{jinglu@pku.edu.cn}
\affiliation{State Key Laboratory for Mesoscopic Physics and School of Physics, Peking University, Beijing 100871, P. R. China}
\affiliation{Collaborative Innovation Center of Quantum Matter, Beijing 100871, P. R. China}
\affiliation{Beijing Key Laboratory for Magnetoelectric Materials and Devices (BKL-MEMD), Peking University, Beijing 100871, P. R. China}
\affiliation{Peking University Yangtze Delta Institute of Optoelectronics, Nantong 226010, P. R. China}
\affiliation{Key Laboratory for the Physics and Chemistry of Nanodevices, Peking University, Beijing 100871, P. R. China}
\affiliation{Beijing Key Laboratory of Quantum Devices, Peking University, Beijing 100871, P. R. China}

\author{Yee Sin Ang}
\email{yeesin\_ang@sutd.edu.sg}
\affiliation{Science, Mathematics and Technology (SMT) Cluster, Singapore University of Technology and Design, Singapore 487372, Republic of Singapore}

\begin{abstract}

Altermagnetism, characterized by zero net magnetization and symmetry-protected spin-split band structures, has recently emerged as a promising platform for spintronics. In altermagnetic tunnel junctions (AMTJs), the suppression of tunneling in the antiparallel configuration relies on the mismatch between spin-polarized conduction channels in momentum space. However, ideal nonoverlapping spin-polarized Fermi surfaces are rarely found in bulk altermagnets. Motivated by the critical influence of Fermi surface geometry on tunneling magnetoresistance (TMR), we investigate three experimentally synthesized altermagnets -- bulk $\mathrm{V_2Te_2O}$, $\mathrm{RbV_2Te_2O}$, and $\mathrm{KV_2Se_2O}$ -- to elucidate how flatband-driven Fermi surfaces minimize spin-channel overlap and boost AMTJ performance. 
Notably, $\mathrm{RbV_2Te_2O}$ and $\mathrm{KV_2Se_2O}$ host flat altermagnetic Fermi sheets, which confine spin degeneracy to minimal arc-like or nodal-like regions. Such Fermi surface geometry drastically reduces spin overlap, resulting in an unprecedented intrinsic TMR well over $10^3\%$ in the $\mathrm{KV_2Se_2O}$-based AMTJ. 
Incorporating an insulating barrier further enhances the TMR to $\sim10^6\%$, surpassing most conventional MTJs. These results not only establish $\mathrm{KV_2Se_2O}$ as a compelling candidate AMTJ material, but also highlight the critical role of flatband Fermi surface geometry in achieving high-performance altermagnetic-spintronic device technology.

\end{abstract}


\maketitle

\section{\label{sec:level1}Introduction}

Magnetoresistive random-access memory (MRAM) is a promising candidate for next-generation nonvolatile memory technology, owing to its fast operation speed and low power consumption \cite{worledge2024spin, jung2022crossbar, sharma2021proposal, chappert2007emergence}. A magnetic tunnel junction (MTJ) forms the core of MRAM, which comprises two ferromagnetic (FM) electrodes separated by an insulating barrier \cite{julliere1975tunneling, jinnai2020scaling, zhang2024electric}.In conventional MTJs, binary information is encoded in the relative magnetization alignment of the FM electrodes, while data writing can be performed electrically through spin-transfer or spin-orbit torques and data reading is typically accomplished through the tunneling magnetoresistance (TMR) effect \cite{vzutic2004spintronics, zheng2025all, liu2025achieving}.
However, FM-based architectures suffer from stray magnetic fields, which limit device density and stability. Antiferromagnetic (AFM) materials inherently eliminate stray fields and exhibit ultrafast spin dynamics in the terahertz regime \cite{shao2024antiferromagnetic, jungwirth2016antiferromagnetic, xiong2022antiferromagnetic}, but their compensated spin structure leads to negligible spin polarization in their band structures, which restricts their potential in spintronics.

Bridging the long-standing dichotomy between ferromagnetism and antiferromagnetism, altermagnetism has emerged as a compelling candidate for MRAM due to its “best-of-both-worlds” character: it features collinear AFM spin ordering together with FM-like spin-split electronic structures \cite{vsmejkal2022emerging, vsmejkal2022beyond, fender2025altermagnetism, Fukaya_2025}. This unique combination enables spin-dependent transport in the absence of net magnetization, offering new opportunities for low-dissipation and high-speed spintronic applications. Unlike conventional antiferromagnets, altermagnets break both $\mathcal{P}\mathcal{T}$ (space inversion and time reversal) and $\mathcal{T}t$ (time reversal combined with half-cell translation) symmetries, leading to sizable nonrelativistic spin splitting away from high-symmetry paths in the Brillouin zone (BZ) \cite{liu2022spin, chen2025unconventional}. These features make altermagnets a compelling building block for MTJs, combining the spin-polarized transport of ferromagnets with the stray-field-free stability characteristic of antiferromagnets \cite{bai2024altermagnetism, shao2024antiferromagnetic, shao2021NC, song2025altermagnets, baltz2018antiferromagnetic}. 

Although multiple altermagnets have been theoretically proposed, their experimental realizations remain limited. Among the reported candidates, RuO$_2$ remains one of the most extensively studied systems, but its altermagnetic nature is still under debate \cite{PhysRevLett.133.176401, noh2025tunneling, fedchenko2024observation, hiraishi2024nonmagnetic, wu2025fermi}. More recently, $\mathrm{KV_2Se_2O}$ \cite{KV2Se2O, sarkar2025altermagnet, wang2025atomicscalespinsensing2d} and $\mathrm{RbV_2Te_2O}$ \cite{zhang2025crystal} have been experimentally confirmed to host $d$-wave altermagnetic band structures through spin- and angle-resolved photoemission spectroscopy (SARPES). Their above-room-temperature N\'eel temperatures and highly anisotropic spin-polarized Fermi surfaces \cite{lai2506d} render them particularly attractive for altermagnet-based spintronic applications.

The transport characteristics and overall performance of an altermagnetic tunnel junction (AMTJ) are critically influenced by the Fermi surface geometry of its electrodes \cite{simon2013oxford, lai2506d}. The extent of spin matching (or mismatching) between spin-polarized conduction channels directly determines the current magnitude in the parallel (P) or antiparallel (AP) state, and hence the tunneling magnetoresistance (TMR) -- a key figure of merit for MRAM technologies. Motivated by the pivotal role of Fermi surface geometry in shaping TMR, we examine three experimentally realized altermagnets -- bulk $\mathrm{V_2Te_2O}$, $\mathrm{RbV_2Te_2O}$, and $\mathrm{KV_2Se_2O}$ -- to elucidate how altermagnetic flatbands can suppress spin channel overlap.

Our first-principles quantum transport simulations show that the quasi-layered altermagnets $\mathrm{RbV_2Te_2O}$ and $\mathrm{KV_2Se_2O}$ host altermagnetic flatbands near the Fermi level, which give rise to strongly anisotropic and quasi-2D Fermi surface sheets. As a direct consequence, intersections between opposite-spin channels are confined to arc-like or nodal-like regions in momentum space, which dramatically reduces the available phase space for spin degeneracy. This flatband-induced suppression of spin channel overlap strongly diminishes the AP-state tunneling, enabling the realization of large TMR ratios. While minimizing spin-channel overlap is a well-established paradigm for achieving large TMR, this work identifies altermagnetic flatbands as a concrete, material-specific mechanism for realizing this condition, thereby providing a viable and materials-guided route to high-performance AMTJs.

In particular, $\mathrm{KV_2Se_2O}$ exhibits only four spin-degenerate, nodal-like overlaps at the Fermi level, yielding an intrinsic TMR as high as $4.3\times10^3\%$ with a vacuum barrier. With an appropriate insulating barrier, the TMR can be further increased to $1.1\times10^6\%$, exceeding the theoretical value of conventional $\mathrm{Fe|MgO|Fe}$ MTJs ($\sim3700\%$) by three orders of magnitude \cite{yuasa2004giant, parkin2004giant}. In contrast, when transport is oriented along a crystallographic direction with larger opposite-spin overlap, the TMR of the $\mathrm{KV_2Se_2O}$ AMTJ drops substantially to $3.3\times10^3\%$. These findings underscore the central role of Fermi surface engineering, particularly through altermagnetic flatband design and crystallographic orientation control, in achieving high-TMR AMTJs. This work not only reveals $\mathrm{KV_2Se_2O}$ as a compelling candidate for AMTJs, but also provides a blueprint for high-performance altermagnetic spintronic architectures, uniquely enabled by the spin-contrasting flatband Fermi surface geometry of quasi-layered altermagnets.

\section{\label{sec:level1}Results}

\subsection{\label{sec:level2}Fermi Surface Geometry in Magnetic Tunnel Junctions}

\begin{figure*}
\centering
\includegraphics[width=0.985\textwidth]{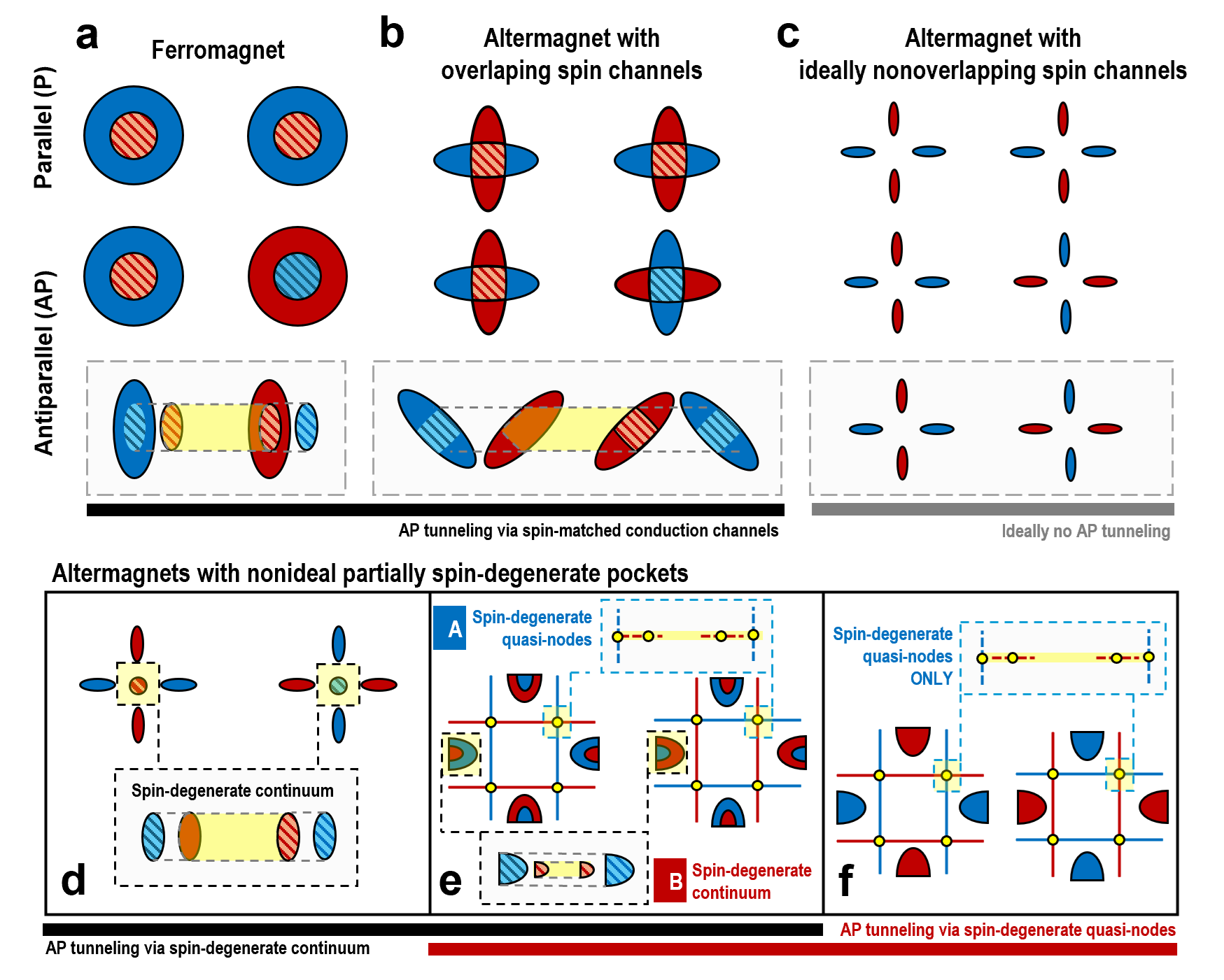}
\caption{\label{fig:epsart} Schematic illustration of tunneling magnetoresistance (TMR) mechanisms in MTJs with different electrode materials. a) Conventional ferromagnets. b) Typical altermagnets with overlapping spin conduction channels. c) AMs with ideally nonoverlapping spin conduction channels. (d-f) AMs with nonideal partially spin-degenerate Fermi pockets that allow residual AP tunneling. Two types of overlap are considered: spin-degenerate continuum and discrete spin-degenerate quasi-nodes. Red and blue denote spin-up and spin-down conduction channels, respectively.}
\end{figure*}

The transport properties of MTJs are governed by spin-dependent quantum tunneling across a thin insulating barrier. The efficiency of this process is strongly dictated by the Fermi surface geometry of the FM electrodes, which determines the momentum-resolved conduction channels available for tunneling (Figure 1) \cite{fu2025light, fu2025floquet}. In the ballistic regime, electron transmission across the barrier requires the conservation of both spin and transverse momentum \cite{datta1997electronic}. Because the Fermi surfaces of the FM electrodes are spin-polarized, the number of available conduction channels differs for majority and minority spins. As a result, the junction resistance depends sensitively on the relative magnetic alignment of the electrodes. When the magnetizations are parallel (P), the majority-spin channels on both sides align, yielding high conductance (low resistance). Conversely, in the antiparallel (AP) configuration, the strong mismatch between majority and minority spin bands significantly reduces the available tunneling pathways, leading to a high-resistance state. This resistance modulation via magnetic alignment leads to a TMR effect along all crystallographic directions in an FM-based MTJ [Figure 1(a)].

In the case of altermagnetic electrodes, strongly anisotropic, spin-polarized Fermi surfaces give rise to momentum-dependent and spin-contrasting conduction channels for opposite spins \cite{chen2025unconventional, fu2506all}. When the N\'eel vectors of the two altermagnetic electrodes are antiparallel, the conduction channels become spin-mismatched, leading to a strong suppression of AP-state tunneling and thus producing a TMR effect \cite{vsmejkal2022giant}. For a typical planar $d$-wave Fermi surface geometry [Figure 1(b)], the rotational symmetry connecting opposite-spin sublattices enforces spin degeneracy at the $\Gamma$ point, yielding a sizable number of spin-degenerate conduction channels. Consequently, AP-state transmission cannot be fully suppressed, and this residual “leakage” current limits the achievable TMR in such AMTJs. 

Ideally, if the spin-resolved Fermi surfaces of the two electrodes exhibit no overlap at any transverse momentum [Figure 1(c)], transmission in AP state can be completely suppressed. C-paired spin–valley locking (SVL) in two-dimensional (2D) lattices provides a symmetry-driven route to realize such well-separated spin-polarized channels \cite{ma2021multifunctional}. However, as interlayer coupling and the restoration of inversion symmetry in multilayer systems tend to hybridize opposite-valley spin states, SVL is typically realized in monolayer systems \cite{xiao2012coupled, schaibley2016valleytronics, xu2014spin, yang2021valley}. This makes direct implementation of SVL in AMTJs challenging, as stable vertical tunneling structures generally require bulk electrodes rather than monolayer components.

\subsection{\label{sec:level2}Candidate Altermangets for Magnetic tunnel Junctions}

As fully separated spin-polarized pockets are rarely realized in three-dimensional (3D) bulk materials, we focus on more experimentally accessible cases of bulk AMs that host partially spin-degenerate conduction channels \cite{zeng2024observation, reimers2024direct}. Achieving large TMR in such systems requires Fermi surface geometries with minimal overlap between opposite-spin conduction channels. In particular, sufficiently flat Fermi surfaces can restrict this overlap to isolated, nodal-like spin-degenerate regions.
Motivated by this consideration, we examine three experimentally synthesized layered altermagnetic candidates -- bulk $\mathrm{V_2Te_2O}$ \cite{doi:10.1021/acs.inorgchem.8b02280}, $\mathrm{RbV_2Te_2O}$ \cite{zhang2025crystal} and $\mathrm{KV_2Se_2O}$ \cite{KV2Se2O} -- each of which hosts highly anisotropic, band-flattened Fermi surfaces. As illustrated in Figures 1(d–f), these systems exhibit distinct forms of residual overlap, ranging from extended spin-degenerate continua to discrete spin-degenerate quasi-nodes. In the following, we systematically investigate their structural and electronic characteristics (see Figure 2) to assess their suitability for AMTJ device applications.

\begin{figure*}
\centering
\includegraphics[width=0.985\textwidth]{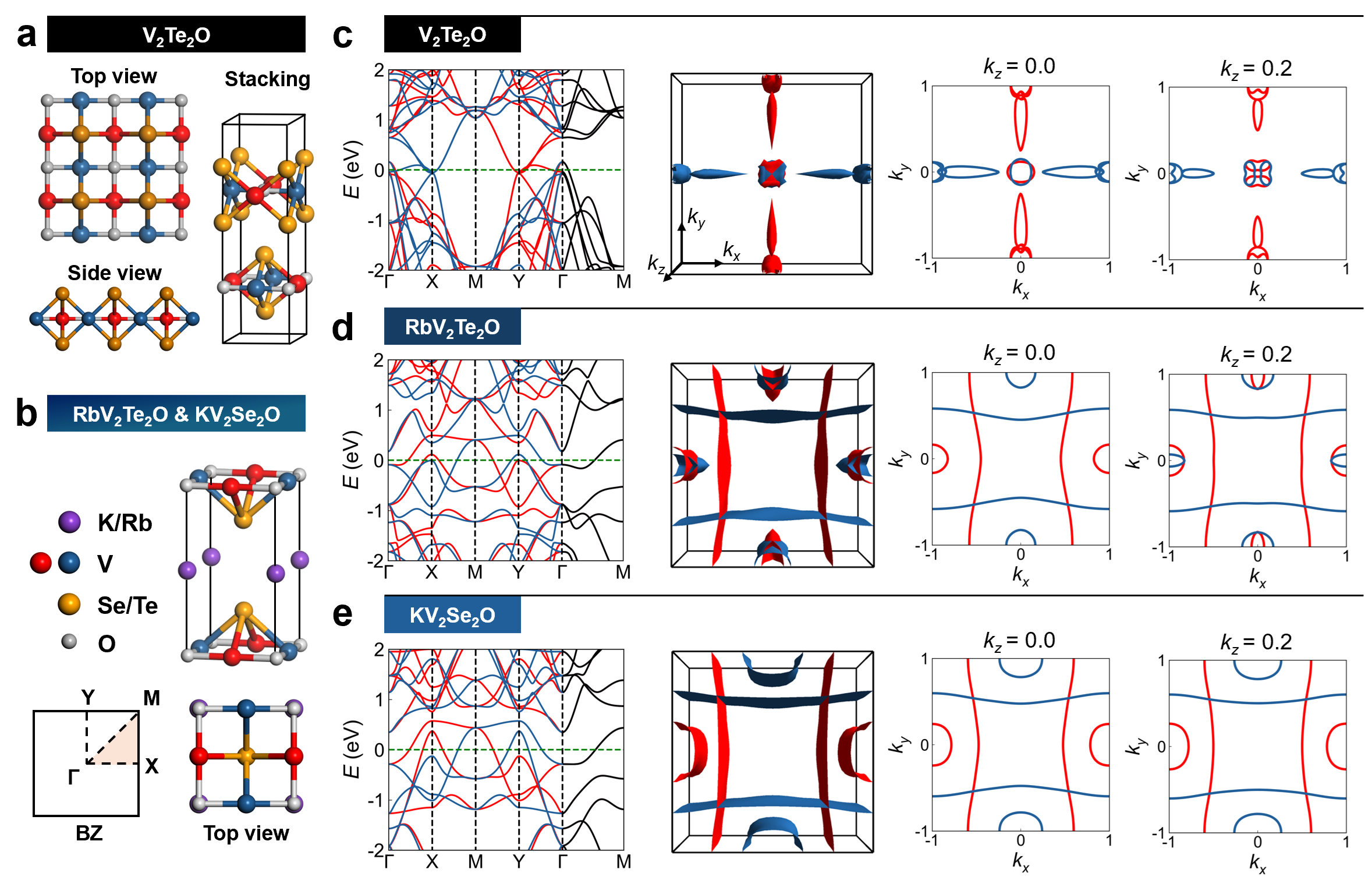}
\caption{\label{fig:epsart} Crystal and magnetic structures of a) $\mathrm{V_2Te_2O}$, b) $\mathrm{RbV_2Te_2O}$ and $\mathrm{KV_2Se_2O}$. Red and blue indicate up and down spins of the V atoms, respectively. (c-e) Electronic band structures, three-dimensional spin-resolved Fermi surfaces and Fermi surface cross-sections at different $k_z$ planes ($k_z=0$ and 0.2) of different materials. Spin-degenerate bands are shown in black.}
\end{figure*}

Monolayer $\mathrm{V_2Te_2O}$ exhibits a tetragonal lattice with a checkerboard configuration [Figure 2(a)]. Each vanadium atom is coordinated by four Te and two O atoms in a distorted octahedral environment, and exhibits staggered out-of-plane magnetization. In the multilayer form, $\mathrm{V_2Te_2O}$ adopts an AB-AFM stacking, where adjacent layers are laterally shifted by half a unit cell and alternate in spin orientations \cite{doi:10.1021/acs.inorgchem.8b02280}. Structurally, the two opposite-spin sublattices are not related by spatial inversion or translation, but are instead connected by $C_{4z}$ or $M_{1\bar{1}0}$ operation. This symmetry arrangement gives rise to a planar $d$-wave altermagnetic order, which is reflected in its spin-split band structures. The energy bands along the $\mathrm{\Gamma-X-M}$ and $\mathrm{\Gamma-Y-M}$ paths exhibit opposite spin splitting [Figure 2(c)], whereas those along the $\mathrm{\Gamma-M}$ path remain spin-degenerate. This momentum-dependent spin texture results in discrete spin-polarized Fermi pockets near the X and Y valleys, whereas multiple band crossings near the $\Gamma$ point generate an extended spin-degenerate continuum at the Fermi level.

$\mathrm{RbV_2Te_2O}$ and $\mathrm{KV_2Se_2O}$ are experimentally confirmed metallic $d$-wave altermagnets \cite{zhang2025crystal, KV2Se2O}. These altermagnets share an identical layered structure, consisting of $\mathrm{V_2Te(Se)_2O}$ sheets separated by Rb/K atoms [Figure 2(b)]. They crystallize in a tetragonal lattice with space group $P_4/mmm$. Nuclear magnetic resonance measurements reveal staggered spin orientations of V atoms along the $c$ axis within each $\mathrm{V_2Te(Se)_2O}$ plane, with a N\'eel temperature well above room temperature. The magnetic structure also breaks both $\mathcal{PT}$ and $\mathcal{T}t$ symmetries, while preserving combined symmetries such as $[C_2||C_{4z}]$ and $[C_2||M_{1\bar{1}0}]$ (the left operation acting in spin space while the right acting in real space). These symmetry characteristics produce spin-split patterns analogous to those of bulk $\mathrm{V_2Te_2O}$ [Figures 2(d,e)]. 
The spin-resolved Fermi surfaces of both compounds consist of quasi-2D sheets extending along $k_z$, with distinct pockets near the Brillouin zone (BZ) boundaries. A comparison of the $k_z = 0.2$ Fermi surface cross-sections reveals that $\mathrm{RbV_2Te_2O}$ retains arc-like regions of residual overlap, while $\mathrm{KV_2Se_2O}$ achieves nearly complete spin separation, with only four spin-degenerate nodal points. The presence of such minimal spin-degenerate regions suggests strong suppression of AP-state leakage current and, consequently, large TMR in these altermagnetic systems.

The emergence of the altermagnetic flatbands observed in these two materials can be traced to a pronounced directional anisotropy in the electronic dispersion \cite{lai2506d} (See Section B of the Supporting information for details). As shown in Figure S3, the band giving rise to the quasi-2D Fermi surface sheets in both $\mathrm{RbV_2Te_2O}$ and $\mathrm{KV_2Se_2O}$ is dominated by the V-$d_{xz}$ orbital for spin-up electrons and by the V-$d_{yz}$ orbital for spin-down electrons. This orbital selectivity induces a strong spin-dependent crystallographic anisotropy between the $x$ and $y$ directions. Moreover, owing to the intercalation of Rb/K atoms, electronic coupling along the $z$ direction is intrinsically weaker than that within the $xy$ plane. As a result, altermagnetic flatbands emerge in these layered compounds, leading to minimal spin channel overlap at the Fermi level.

\subsection{\label{sec:level2}Intrinsic transport properties of V$_2$Te$_2$O, RbV$_2$Te$_2$O and KV$_2$Se$_2$O AMTJs}

\begin{figure*}
\centering
\includegraphics[width=0.985\textwidth]{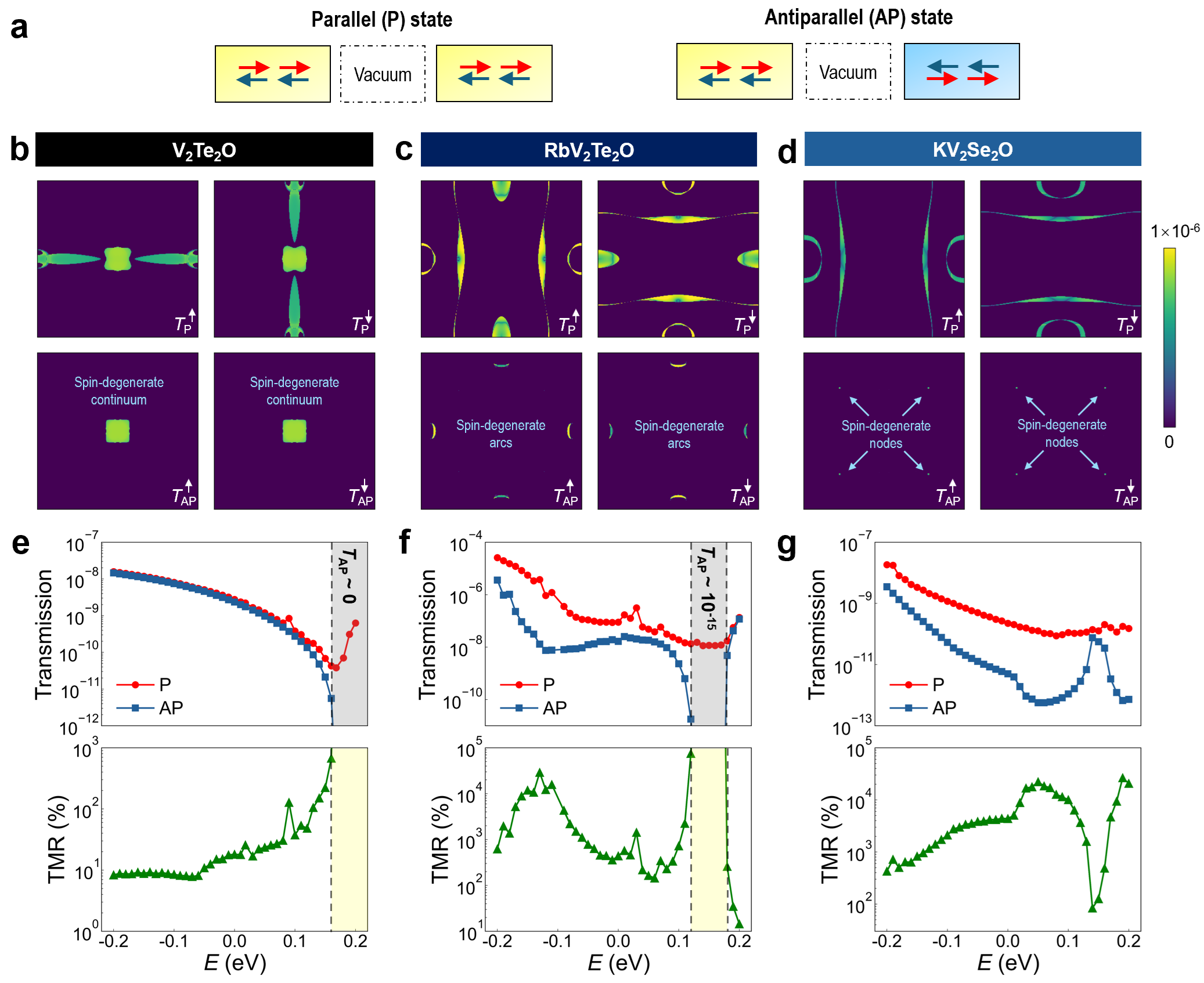}
\caption{\label{fig:epsart} Intrinsic transport properties of AMTJs with vacuum barriers based on $\mathrm{V_2Te_2O}$, $\mathrm{RbV_2Te_2O}$ and $\mathrm{KV_2Se_2O}$ electrodes. a) Schematic illustration of magnetization alignments in the parallel (P) and antiparallel (AP) configurations. (b-d) Calculated spin-resolved, $k_\parallel$-resolved transmission coefficients at the Fermi level for each AMTJ in both P and AP states. (e-g) Energy-dependent total transmission spectra for P and AP states, along with the resulting TMR ratio. Shaded regions highlight the energy ranges where AP tunneling $T_\mathrm{AP}$ is significantly suppressed due to changes in available conduction channels when energy is away from the Fermi level. The normalized TMR ratios, defined as $(T_\mathrm{P}-T_\mathrm{AP})/T_\mathrm{P}$, are provided in Figure S4 of the Supporting Information.}
\end{figure*}

Having established the Fermi surface geometries of the three representative bulk altermagnets, we now turn to their intrinsic transport properties along the [001] direction. As a preliminary assessment, the AMTJs are modeled using an ideal 6\AA vacuum spacer [Figure 3(a)]. Such modelling setup enables a more direct assessment on how Fermi surface geometry inherent to each material influences the resulting TMR performance, before addressing the complicated interfacial effects when an insulating barrier is employed. Figures 3(b-d) show the transverse-momentum ($k_\parallel$)-resolved transmission coefficients at the Fermi level for spin-up and spin-down electrons. In the P state, the transmission profiles closely resemble the projections of the anisotropic Fermi surfaces onto the [001] plane, as the spin-matched bands of the two electrodes generate efficient conduction channels. In contrast, electron transmission in the AP state is strongly suppressed throughout the BZ, except in several regions where spins are degenerate, namely (i) an extended continuum for $\mathrm{V_2Te_2O}$ [Figure 3(b)]; (ii) arc-like features for $\mathrm{RbV_2Te_2O}$ [Figure 3(c)]; and (iii) nodal-like features for $\mathrm{KV_2Se_2O}$ [Figure 3(d)]. 

The $\mathrm{V_2Te_2O}$-based AMTJ exhibits pronounced transmission features for both spin channels in the P state, as characterized by a broad high-intensity region centered at the $\Gamma$ point and extended continua along the -$\mathrm{X}-\Gamma-\mathrm{X}$ or -$\mathrm{Y}-\Gamma-\mathrm{Y}$ direction. The spin-up and spin-down transmission patterns share identical geometries, differing only by a $\pi/2$ rotation, which is consistent with the $d$-wave symmetry of the altermagnetic order. Consequently, the total current along the [001] direction remains effectively spin-neutral. Owing to the large spin-degenerate continuum around the $\Gamma$ point, AP-state tunneling cannot be completely suppressed, giving rise to a residual leakage current. 

The evolution of the total transmission with energy further reveals the underlying mechanism [Figure 3(e)]. In both configurations, the transmission initially decreases with increasing energy. Beyond $E > 0.16$ eV, however, the total transmission in the P state $T_\mathrm{P}$ ($T_\mathrm{P}=T_\mathrm{P}^\uparrow+T_\mathrm{P}^\downarrow$) exhibits a sharp increase while $T_\mathrm{AP}$ drops abruptly to nearly zero (shaded region). This sudden contrast originates from the evolution of the bulk $\mathrm{V_2Te_2O}$ Fermi surfaces at higher energies. As illustrated in Figure 4, increasing the energy from 0.06 to 0.18 eV progressively reduces the spin-degenerate continuum near the $\Gamma$ point, which eventually disappears at $E = 0.18$ eV. At this energy, $\mathrm{V_2Te_2O}$ becomes an ideal electrode material for AMTJs, exhibiting perfectly separated spin-polarized conduction channels. The energy-resolved TMR ratio shown in Figure 3(e) is defined as $\mathrm{TMR}=(T_\mathrm{P}-T_\mathrm{AP})/T_\mathrm{AP}$. 
Owing to the substantial spin channel overlap around the $\Gamma$ point, the $\mathrm{V_2Te_2O}$-based AMTJ yields a modest TMR of only 18\% at $E_\mathrm{F}$.

\begin{figure}
\centering
\includegraphics[width=0.4\textwidth]{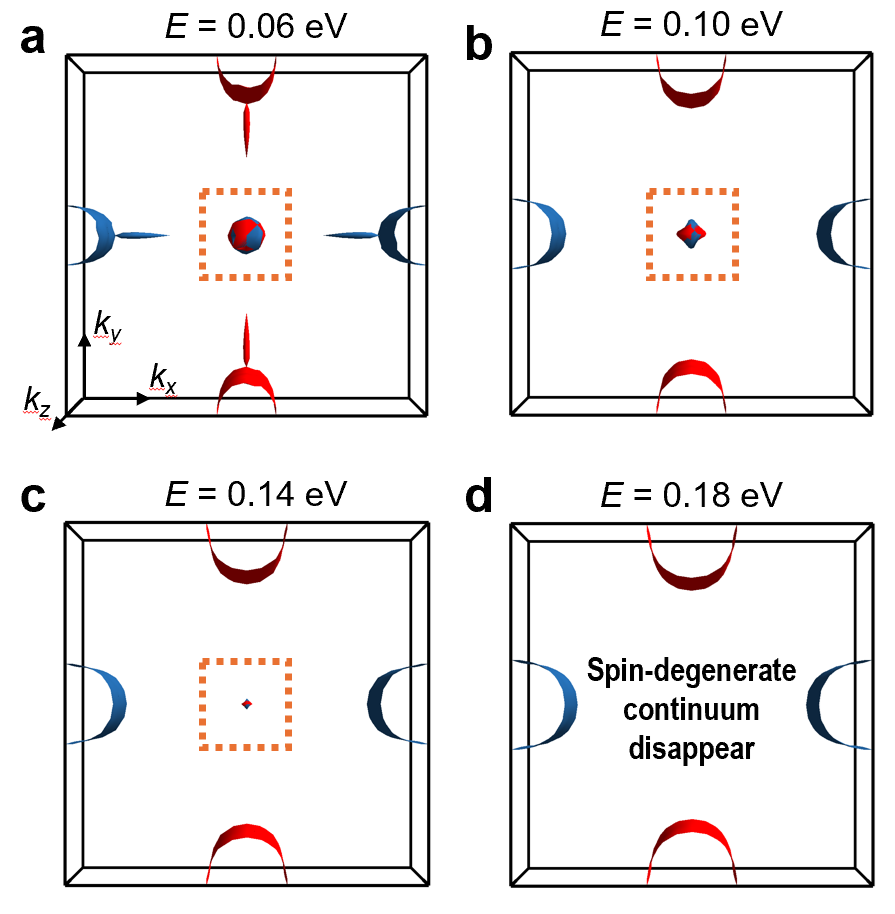}
\caption{\label{fig:epsart} Three-dimensional Fermi surfaces of bulk $\mathrm{V_2Te_2O}$ at a) $E=0.06$ eV, b) $E=0.10$ eV, c) $E=0.14$ eV and d) $E=0.18$ eV. Red and blue indicate spin-up and spin-down Fermi surfaces, respectively. The spin-degenerate continuum near the $\Gamma$ point progressively diminishes with increasing energy and entirely vanishes at $E = 0.18$ eV.}
\end{figure}

In contrast to $\mathrm{V_2Te_2O}$, the Fermi surface geometries of $\mathrm{RbV_2Te_2O}$ and $\mathrm{KV_2Se_2O}$ exhibit no overlap near the $\Gamma$ point [Figures 3(c,d)]. Both AMTJs feature narrow transmission channels in the P state, with distinct Fermi pockets at the BZ boundary, which is consistent with their quasi-2D altermagnetic nature. However, the spin-polarized pockets in $\mathrm{RbV_2Te_2O}$ are not completely separated, giving rise to several extended spin-degenerate “arcs" that sustain AP transmission. By contrast, the $\mathrm{KV_2Se_2O}$-based AMTJ hosts only four discrete, “nodal-like" intersection points between opposite-spin conduction channels. These isolated nodes, providing minimal conduction pathway, immediately suggest a strong suppression of AP-state tunneling in the device.

The distinction between the spin-degenerate “arcs” in $\mathrm{RbV_2Te_2O}$ and the discrete “nodal points” in $\mathrm{KV_2Se_2O}$ is quantitatively reflected in their total transmission spectra and the corresponding TMR ratios [Figure 3(f,g)]. In $\mathrm{RbV_2Te_2O}$, the persistence of spin-degenerate “arcs" leads to a moderate TMR ratio of 435\% at the Fermi level. In contrast, the $\mathrm{KV_2Se_2O}$-based AMTJ exhibits a pronounced difference between the P- and AP-state transmissions at $E_\mathrm{F}$, resulting in a much higher TMR of $4.3\times10^3\%$. The shaded region in Figure 3(f) highlights a reduction in AP tunneling for $\mathrm{RbV_2Te_2O}$, which can be attributed to subtle changes in its Fermi surface geometry (see Section D of the Supporting Information for details). 

Despite the broader spin channel overlap in $\mathrm{V_2Te_2O}$, the absolute values of the AP-state transmission coefficients at specific $k$-points are significantly larger in the $\mathrm{RbV_2Te_2O}$ device [Figure 3(c)]. For example, $T_\mathrm{AP}^\uparrow$ near the $\Gamma$ point reaches $\sim 10^{-7}$ in the $\mathrm{V_2Te_2O}$ AMTJ, whereas $T_\mathrm{AP}^\uparrow$ at the arc-like features along the $\Gamma$–X direction in the $\mathrm{RbV_2Te_2O}$ device is on the order of $10^{-5}$. This pronounced disparity in the magnitude of the transmission coefficients ultimately leads to a larger total AP-state transmission at the Fermi level in the $\mathrm{RbV_2Te_2O}$ MTJ.

Taken together, the contrasting forms of the spin-degenerate regions -- continua in $\mathrm{V_2Te_2O}$, arcs in $\mathrm{RbV_2Te_2O}$, and discrete nodal points in $\mathrm{KV_2Se_2O}$ -- together with their corresponding order-of-magnitude variations in the TMR, underscore the pivotal role of Fermi-surface overlap in determining and pushing the performance limits of AMTJs.

\subsection{\label{sec:level2}Tunneling magnetoresistance of KV$_2$Se$_2$O-based AMTJ with insulating spacers}

\begin{figure*}
\centering
\includegraphics[width=0.985\textwidth]{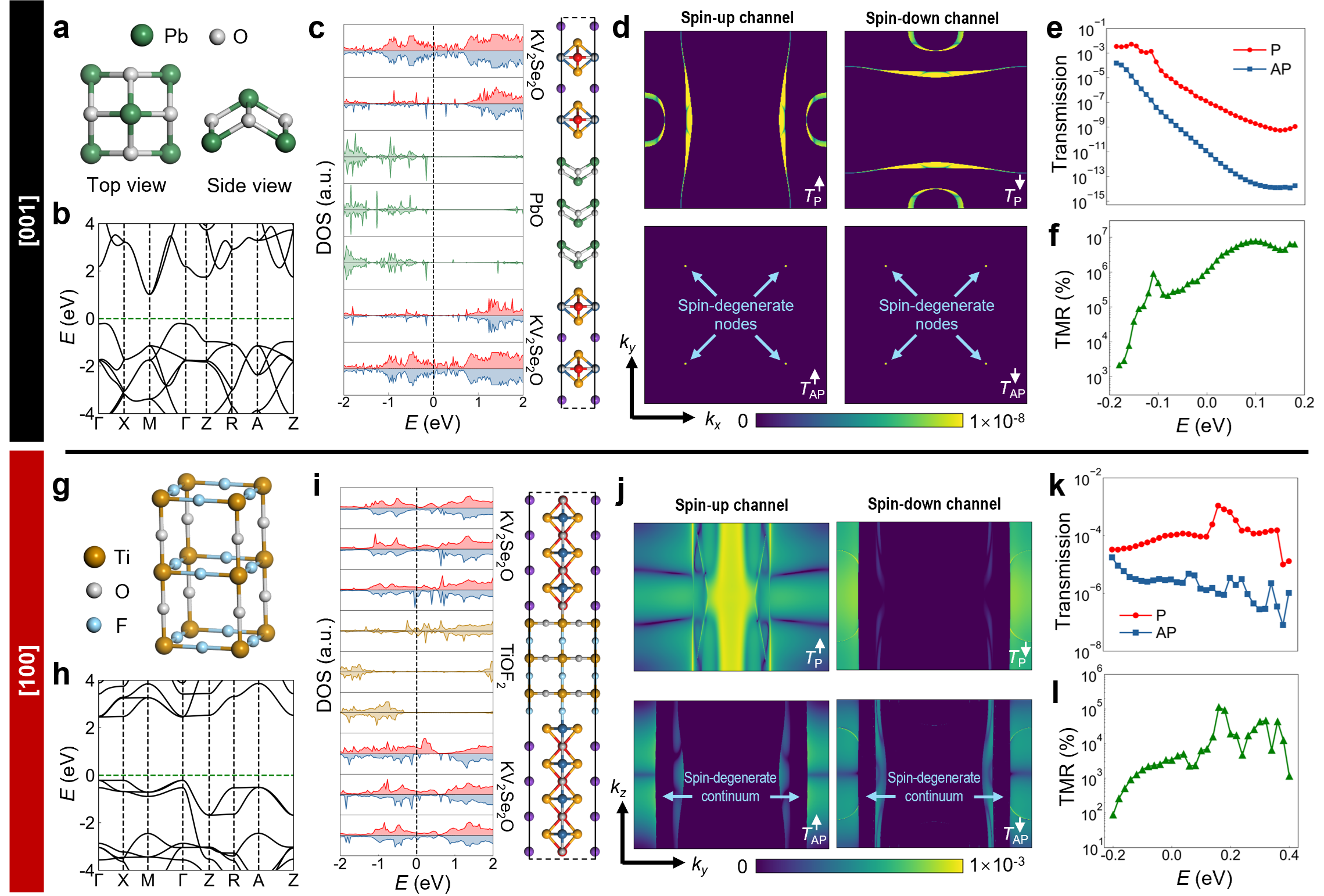}
\caption{\label{fig:epsart} Transport properties of $\mathrm{KV_2Se_2O}$-based AMTJs along the (a-f) [001] and (g-i) [100] directions. a) Crystal structure of the barrier material PbO. b) Electronic band structures of bulk PbO. c) Atomic structure and layer-resolved density of states (LDOS) of the $\mathrm{KV_2Se_2O}|\mathrm{PbO}|\mathrm{KV_2Se_2O}~(001)$ MTJ. Each panel corresponds to a single atomic layer of either $\mathrm{KV_2Se_2O}$ or PbO, with spin-up and spin-down components shown in the upper and lower subpanels. The Fermi level is marked by a dashed line. d) $k_\parallel$-resolved spin-polarized transmission coefficients at the Fermi level for the (001) AMTJ in both P and AP states. e) Total transmission spectra and f) TMR ratio as functions of energy. (g-l) Same as (a-f) but for the $\mathrm{KV_2Se_2O}|\mathrm{TiOF_2}|\mathrm{KV_2Se_2O}$ MTJ along the [100] direction.}
\end{figure*}

Having identified $\mathrm{KV_2Se_2O}$ as the most promising electrode material with intrinsically high TMR at the Fermi level, we next investigate its performance in more realistic device architectures that incorporate an insulating barrier. The choice of barrier can critically influence both tunneling efficiency and spin selectivity, as its electronic structure and symmetry compatibility with the electrodes directly affect the available tunneling pathways \cite{PhysRevB.69.174408}. An appropriately chosen barrier can therefore substantially enhance the achievable TMR ratio beyond the intrinsic limit imposed by the electrode Fermi surface. To this end, we construct $\mathrm{KV_2Se_2O}$-based AMTJs incorporating realistic barrier materials along the [001] and [100] directions, and systematically evaluate their transport characteristics.

For transport along the [001] direction, we select PbO as the tunnel barrier owing to its excellent structural compatibility with $\mathrm{KV_2Se_2O}$ and a small lattice mismatch of only 1.1\%. PbO is a well-established van der Waals semiconductor with a tetragonal crystal structure \cite{zhu2025altermagnetic}, as illustrated in Figure 5(a). A monolayer of PbO consists of an oxygen atomic layer sandwiched between two lead layers arranged in a checkerboard lattice. In multilayer form, it adopts an AA-stacking configuration, in which adjacent PbO layers are aligned vertically with an interlayer spacing of 4.97 \AA.
Owing to the absence of magnetic elements in its unit cell, bulk PbO exhibits spin-degenerate electronic bands and a band gap of 1.23 eV [Figure 5(b)]. In our AMTJ design, a three-layer PbO barrier is employed to separate the two electrodes. The layer-resolved density of states (LDOS) [Figure 5(c)] reveals a well-defined band gap at the Fermi level within the barrier region, confirming that the PbO barrier effectively suppresses direct electronic coupling between the electrodes and ensures transport dominated by quantum tunneling.

The $k_\parallel$-resolved transmission coefficients for the $\mathrm{KV_2Se_2O}$-based AMTJ along the [001] direction are shown in Figure 5(d). The transmission patterns in the P state closely resemble those of the intrinsic (vacuum-barrier) case, while the nodal-like features remain well preserved in the AP configuration. The total transmission in the P state is enhanced by nearly three orders of magnitude compared with the vacuum-barrier junction [Figure 5(e)], whereas the increase in the AP-state transmission is much smaller. Specifically, $T_\mathrm{P}$ rises from $2.2\times10^{-10}$ to $1.4\times10^{-7}$, while $T_\mathrm{AP}$ increases only from $5.0\times10^{-12}$ to $1.2\times10^{-11}$ at the Fermi level. Consequently, a giant TMR ratio of $1.1\times10^{6}\%$ is achieved at $E_\mathrm{F}$, exceeding the previously reported value of $\sim 500\%$ in $\mathrm{RuO_2}|\mathrm{TiO_2}|\mathrm{RuO_2}$ AMTJs \cite{shao2021NC} by more than three orders of magnitude.

The enhancement of TMR upon introducing an appropriate tunnel barrier can be understood within the framework of tunneling theory \cite{heine1965theory}. Electron transmission through an insulating barrier decays exponentially as $e^{-\kappa d}$, where $d$ denotes the barrier thickness and $\kappa$ is the decay rate determined by the complex band structure of the barrier material \cite{mavropoulos2000complex}. As shown in Figure S6, the lowest decay-rate distribution in the band gap of PbO develops extended low-$\kappa$ regions at larger in-plane momenta $k_\parallel$, which closely coincide with the momentum-resolved conduction channels of the $\mathrm{KV_2Se_2O}$ electrodes. In contrast to the vacuum-barrier case, where the decay rate is nearly independent of $k_\parallel$, this momentum-selective matching substantially enhances the total transmission in the parallel configuration, thereby leading to a pronounced increase in the TMR ratio.

Building on the momentum-selective tunneling mechanism discussed above, the use of an altermagnetic semiconductor as the tunnel barrier provides an additional route to further enhance the TMR via spin-dependent filtering effects \cite{lukashev2012spin, butler2001spin}. Unlike nonmagnetic semiconductors, altermagnetic semiconductors possess symmetry-enforced spin-split band structures, giving rise to evanescent states with different decay rates for spin-up and spin-down electrons \cite{samanta2025spin}. With appropriate matching between the barrier and the electrodes, this spin-dependent decay can provide an additional contribution to the TMR effect \cite{chi2025anisotropic, yang2025unconventional}. As discussed in Section F of the Supporting Information, we employ multilayer $\mathrm{Cr_2Se_2O}$ as the tunnel barrier, which is a van der Waals $d$-wave altermagnetic semiconductor sharing the same crystal structure as $\mathrm{V_2Te_2O}$. Owing to the van der Waals nature of the interface between $\mathrm{Cr_2Se_2O}$ and $\mathrm{KV_2Se_2O}$, the magnetic coupling across the interface is expected to be weak, thereby ensuring device stability \cite{wang2025large}. However, due to the relatively small band gap (0.26 eV) of bulk $\mathrm{Cr_2Se_2O}$, electronic states from the metallic electrodes penetrate more strongly into the barrier region, resulting in enhanced leakage in the AP state and, consequently, a substantially reduced TMR of $9.1\times10^{4}\%$.

\begin{figure*}
\centering
\includegraphics[width=0.9\textwidth]{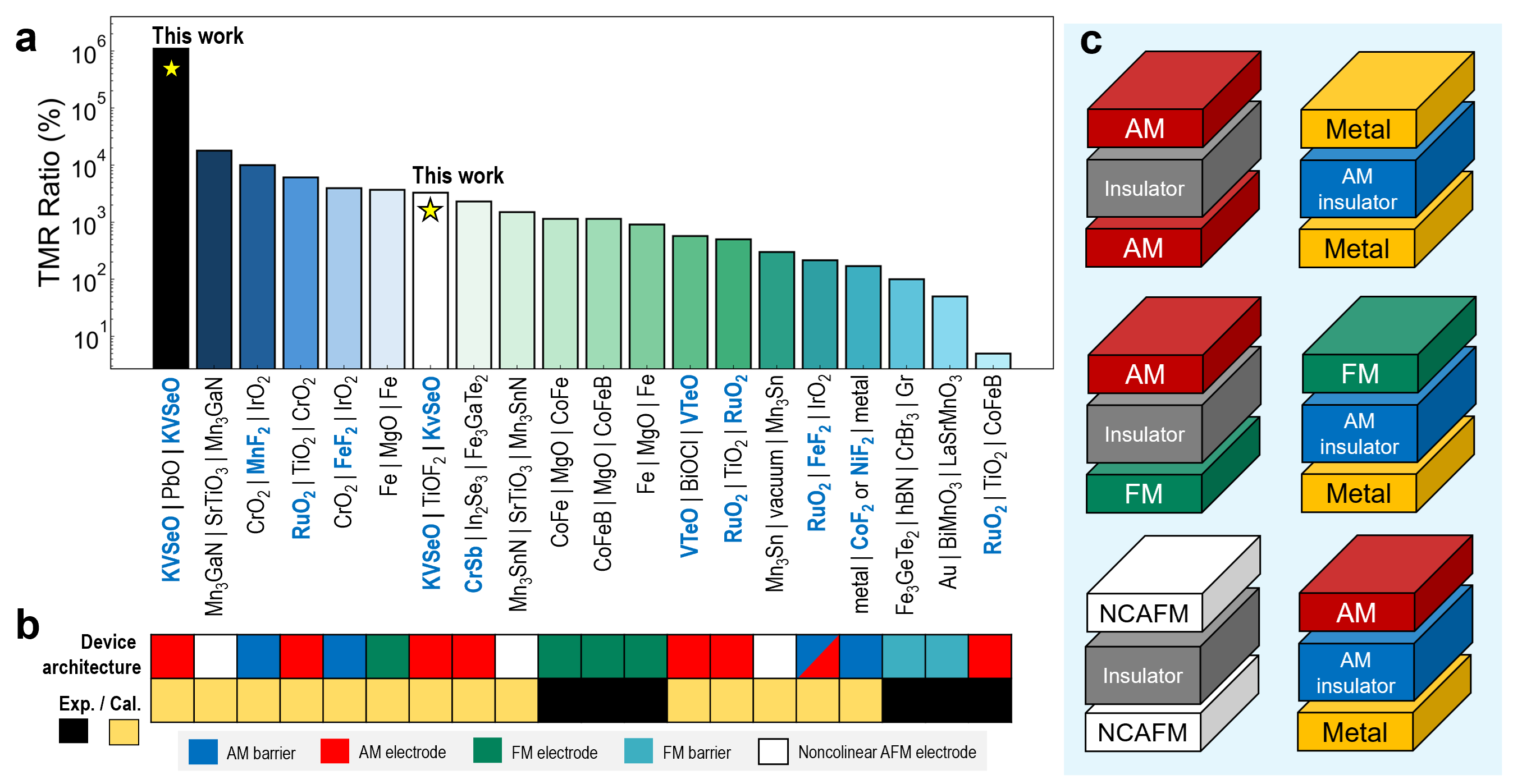}
\caption{\label{fig:epsart} Benchmark of MTJ performances. a) Comparison of TMR ratios obtained using $\mathrm{KV_2Se_2O}$ electrodes in this work (highlighted with stars) and selected results from previous experimental and theoretical studies \cite{gurung2024q, liu2024giant, samanta2024tunneling, chi2025anisotropic, waldron2006first, zhang2025above, liu2025mn3snn, chi2024crystal, cui2023giant, shao2021NC, yuasa2004giant, parkin2004giant, dong2022tunneling, samanta2025spin, noh2025tunneling, scheike2021exceeding, ikeda2008tunnel, scheike2023631}. To ensure a consistent comparison, all experimental values listed here are measured at low temperatures. Altermagnetic materials are marked with blue. For simplicity, $\mathrm{KV_2Se_2O}$ and $\mathrm{V_2Te_2O}$ are abbreviated as KVSeO and VTeO, respectively and "Gr" denotes graphene. b) Device architectures and methodology (experimental or computational) corresponding to each bar in panel (a). The top row codes the type of functional layer -- AM barrier, AM electrode, FM electrode, FM barrier or noncollinear AFM (NCAFM) electrode, while the bottom row distinguishes between experiment (Exp.) and calculation (Cal.). c) Schematic illustrations of representative MTJ architectures with AM or NCAFM.}
\end{figure*}

\subsection{\label{sec:level2}Crystal facet engineering in KV$_2$Se$_2$-based AMTJ}

Due to the pronounced anisotropy of the Fermi surface in $\mathrm{KV_2Se_2O}$, its transport characteristics exhibit strong directional dependence. When the current flows along the [100] direction, the symmetry constraints ($[C_2||C_\mathrm{4z}]$) responsible for spin neutrality in the [001] orientation are lifted, resulting in spin-polarized transport analogous to that in conventional FM-based MTJs. To ensure structural compatibility and minimize lattice strain, the semiconductor $\mathrm{TiOF_2}$ is chosen as the insulating barrier. The atomic and electronic structures of $\mathrm{TiOF_2}$ are shown in Figures 5(g,h). $\mathrm{TiOF_2}$ crystallizes in a tetragonal lattice and exhibits an excellent lattice match with $\mathrm{KV_2Se_2O}$ (mismatch $<1\%$). Furthermore, $\mathrm{TiOF_2}$ possesses a wide band gap of 2.67 eV, which remains well preserved in the $\mathrm{KV_2Se_2O}|\mathrm{TiOF_2}|\mathrm{KV_2Se_2O}$ (100) heterostructure [Figure 5(i)].

The transmission along the [100] direction [Figure 5(f)] exhibit distinct patterns compared with that along [001]. For spin-up electrons, the \textit{quasi-2D} Fermi surface sheets extended along $k_\mathrm{z}-k_\mathrm{y}$ plane form continuous conduction channels throughout the BZ, resulting in a relatively high transmission. In contrast, the spin-down Fermi sheets, oriented primarily along the $k_\mathrm{z}$–$k_\mathrm{x}$ plane, have only limited projection onto the [100] transport plane. Consequently, $T_\mathrm{P}^{\downarrow}$ is dominated by isolated spin-down Fermi pockets located near $k_\mathrm{y} = \pm \pi/a$. In the AP configuration, the number of available transmission states is markedly reduced, with tunneling restricted to small regions near the Brillouin zone boundaries where the spin-down pockets overlap with the opposite-spin Fermi sheets. The resulting spin polarization, defined as $p = (T^\uparrow - T^\downarrow)/(T^\uparrow + T^\downarrow)$, reaches 73.2\% when the N\'eel vectors of the $\mathrm{KV_2Se_2O}$ electrodes are aligned.

By integrating over $k_\parallel$, we obtain the total transmission spectra and corresponding TMR ratio as functions of energy [Figures 5(k,l)]. Because both the P and AP configurations host a larger number of available conduction channels, the overall transmission along the [100] direction is substantially enhanced. At the Fermi energy, the TMR reaches $3.3 \times 10^3\%$, which is much smaller than that of the [001] device due to the spin-degenerate continua that yield sizable AP-state tunneling. Here, [100] configurations have the advantage of overall larger transmission value in the P states, hence better signal-to-noise ratio during device readout \cite{manipatruni2019scalable}, but this comes at the cost of a reduced TMR relative to the [001] counterpart. The pronounced contrast between the [001] and [100] configurations thus underscores the crystal facet engineering in tailoring the performance metrics of AMTJs.

\subsection{Performance Benchmarking}

We provide a performance benchmark of $\mathrm{KV_2Se_2O}$-based AMTJs in Figure 6. Their TMR ratios are compared with representative experimental and theoretical results from recent MTJ studies [Figs. 6a,b; see Section F of the Supporting Information for details] under different device configurations [see Figure 6(c) for representative MTJ setups]. We note that MTJs based on multilayer $\mathrm{CrI_3}$ tunnel barriers, although reported to exhibit giant TMR, rely on a magnetic-field–induced phase transition of the barrier \cite{song2018giant, kim2018one, wang2018very} and its device operation is restricted to cryogenic temperatures due to the low critical temperature of $\mathrm{CrI_3}$ \cite{huang2017layer}. Owing to this fundamentally different mechanism and limited operating temperature window, such devices are not included in our benchmark. 

Under typical MTJ operating conditions with low applied bias voltages (on the order of $10^{-1}$ V), the transport characteristics are primarily governed by the transmission characteristics at the Fermi level \cite{he2023sensitivity,liu2015manipulation}. Therefore, for quantitative benchmarking and direct comparison, we focus on the TMR values evaluated at $E_\mathrm{F}$ for simulation results.
Conventional FM MTJs, such as Fe$|$MgO$|$Fe, typically yield theoretical TMR values around 3700\% \cite{waldron2006first}, whereas the highest experimentally reported TMR values have reached 417\% at room temperature (RT) and 914\% at 3 K \cite{scheike2021exceeding}. For CoFe$|$MgO$|$CoFe or CoFeB$|$MgO$|$CoFeB devices, higher TMR ratios have been achieved recently, exceeding 600\% at RT and 1100\% at low temperatures \cite{ikeda2008tunnel, scheike2023631}. In comparison, the simulated TMR ratios of $\mathrm{RuO_2}$-based AMTJs remain on the order of $10^3\%$ \cite{shao2021NC, samanta2024tunneling, chi2024crystal}. 

Notably, our proposed $\mathrm{KV_2Se_2O|PbO|KV_2Se_2O}$ [001] device achieves a record-high TMR of $1.1\times10^6\%$, surpassing all previously reported AMTJs here by more than two orders of magnitude. This exceptional performance originates from the combination of: (i) the altermagnetic flatband Fermi surface geometry of $\mathrm{KV_2Se_2O}$ which yields minimal nodal-like conduction channels in the AP state; combined with (ii) the symmetry-matched PbO insulating barrier. When transport occurs along the [100] direction, the TMR is substantially reduced due to the presence of extended spin-matched conduction channels in the AP configuration -- though the resulting TMR remains comparable to many other MTJ designs. Overall, our computational results establish the theoretical performance ceiling of $\mathrm{KV_2Se_2O}$-based AMTJs, providing both a quantitative benchmark for material selection and a conceptual framework for the Fermi-surface–engineered design of next-generation spintronic devices.

\section{\label{sec:level2}Conclusion}

In summary, we have systematically investigated the intrinsic transport characteristics in AMTJs using three experimentally realized bulk altermagnets -- $\mathrm{V_2Te_2O}$, $\mathrm{RbV_2Te_2O}$, and $\mathrm{KV_2Se_2O}$. We demonstrated that the degree of overlap between opposite-spin conduction channels serves as the key factor governing the TMR performance. Among these materials, $\mathrm{KV_2Se_2O}$ exhibits nearly nodal-like spin-degenerate conduction channels in the AP state, which arises from its altermagnetic flatband Fermi surface geometry. As a result, an intrinsic TMR of $4.3\times10^3\%$ was obtained, which was further enhanced to $1.1\times10^6\%$ by employing a symmetry-matched $\mathrm{PbO}$ insulating barrier in an AMTJ device setup. The pronounced contrast between the [001] and [100] transport directions underscores the essential role of Fermi surface geometry engineering via crystallographic orientation, which enables the overall magnitude of the transmission and the TMR to be tuned and balanced depending on the application needs. 

Beyond identifying $\mathrm{KV_2Se_2O}$ as a compelling altermagnetic electrode material, our results revealed the pivotal role of altermagnetic flatbands in achieving ultrahigh-TMR AMTJs with minimal spin-degenerate “arc-" or “node-" like conduction channels.
More broadly, this work established a concrete design principle for altermagnetic-spintronic devices: high-performance AMTJs are enabled when quasi-layered altermagnets hosting flatband-driven, symmetry-protected Fermi surface geometries that minimize opposite-spin overlap. This flatband-centric paradigm provides a powerful materials design and screening guideline and paves a pathway toward next-generation altermagnetic-spintronics uniquely empowered by quasi-layered altermagnets.

\section{\label{sec:level2}Experimental Section}

\subsection{Density Functional Theory Simulations}

The geometry optimization and electronic structure calculations for $\mathrm{V_2Te_2O}$, $\mathrm{RbV_2Te_2O}$ and $\mathrm{KV_2Se_2O}$ are carried out using the Vienna Ab initio Simulation Package (VASP) \cite{kresse1993ab,kresse1996efficient}. The projector augmented wave (PAW) method is employed, with the exchange-correlation effects treated within the generalized gradient approximation (GGA) using the Perdew–Burke–Ernzerhof (PBE) functional \cite{perdew1996generalized}. We set the plane-wave cutoff energy to 600 eV and the atomic positions are relaxed until the residual forces on each atom are less than $10^{-2}$ eV/\AA. The Brillouin zone is sampled using dense Monkhorst–Pack $k$-point grids of $25\times25\times13$ for bulk material unit-cell calculations. For AMTJ device optimizations, $23\times23\times1$ and $23\times13\times1$ meshes are employed for transport along the [001] and [100] directions, respectively. The on-site Coulomb interaction is treated within the GGA+$U$ scheme \cite{dudarev1998electron, anisimov1991band}, with $U_\mathrm{eff}=3.7$ eV on V 3$d$ orbitals for $\mathrm{V_2Te_2O}$ \cite{doi:10.1021/acs.inorgchem.8b02280} and $1.0$ eV for $\mathrm{RbV_2Te_2O}$ \cite{zhang2025crystal}. We do not apply a Hubbard $U$ correction when calculating $\mathrm{KV_2Se_2O}$, as our tests [see Section A of the Supporting Information for details] show that this setting yields band structures most consistent with the experimental SARPES results \cite{KV2Se2O}. To minimize lattice mismatch and the associated structural distortion, we employ PbO and $\mathrm{TiOF_2}$ as the tunnel barriers for the $\mathrm{KV_2Se_2O}$ AMTJ along the [001] and [100] directions, respectively and we apply a Hubbard $U$ correction of $4.2$ eV on Ti 3$d$ orbitals \cite{legein2024correlated}. 

\subsection{Quantum Atomistic Device Simulations}

The spin-dependent transport properties are evaluated using Density Functional Theory (DFT) combined with the nonequilibrium Green’s function (NEGF) formalism, implemented in the QuantumATK package \cite{smidstrup2019quantumatk, brandbyge2002density}. Our device is modeled as a two-probe system, consisting of a central region and two electrodes. The spin-resolved transmission coefficient $T_{k_{\parallel}, s}(E)$ can be written as
\begin{equation*}
T_{k_{\parallel}, s}(E) = \mathrm{Tr}[G_{k_{\parallel}, s}(E) \Gamma^{l}_{k_{\parallel}, s}(E) G_{k_{\parallel}, s}(E)^\dagger \Gamma^{r}_{k_{\parallel}, s}(E)]
\end{equation*}
where $k_{\parallel}$ denotes the reciprocal lattice vector parallel to the electrode interface (orthogonal to the transport direction) in the irreducible Brillouin zone, and $s$ labels the spin. Here, $G_{k_{\parallel},s}(E)$ is the retarded Green’s function for spin $s$ at $k_{\parallel}$, while $\Gamma^{l(r)}_{k_{\parallel},s}(E)$ denotes the broadening function of the left (right) electrode, which equals the imaginary part of the corresponding self-energy term \cite{williams1982green}. The AMTJ simulations employ a density mesh cutoff of 180 Hartree and an electronic temperature of 300 K. For the self-consistent calculations, the Brillouin zone is sampled using $k$-point meshes of $23\times23\times1$ and $23\times13\times1$ for devices oriented along the [001] and [100] directions, respectively. For the transmission calculations, much denser $k$-point meshes of $301\times301$ and $201\times101$ are used to reliably resolve the minimal spin-channel overlap.
To accurately describe the van der Waals interaction in $\mathrm{V_2Te_2O}$ multilayers, Grimme's DFT-D3 correction is applied \cite{grimme2006semiempirical}.

\subsection{Statistical Analysis}

The results presented in this work are obtained from first-principles electronic structure and quantum transport calculations. No statistical methods or hypothesis testing were employed, as the calculations do not involve stochastic sampling or experimental variability. All data are presented as direct computational outputs without preprocessing, normalization, or outlier treatment. Sample size, measures of variance, and statistical significance testing are therefore not applicable. All simulations and analyses were carried out using VASP, QuantumATK and post-processed with Python-based analysis tools.

\begin{acknowledgements}
This work is supported by the National Natural Science Foundation of China (No. 12274002 and 91964101), the Ministry of Science and Technology of China (No. 2022YFA1203904), the Fundamental Research Funds for the Central Universities, and the High-performance Computing Platform of Peking University. Y.S.A. acknowledges the supports from the Kwan Im Thong Hood Cho Temple Early Career Chair Professorship in Sustainabiltiy, and the Ministry of Education, Singapore, Academic Research Fund Tier 2 under the award number MOE-T2EP50224-0021.
\end{acknowledgements}

\bibliography{Reference}

\end{document}